# Pitch tuning induced by optical torque in heliconical cholesteric liquid crystals

G. Nava[1], F. Ciciulla[1], O. S. Iadlovska[2], O. D. Lavrentovich[2], F. Simoni[1] and L. Lucchetti[1]

1. Università Politecnica delle Marche, Dipartimento SIMAU, via Brecce Bianche, 60131 Ancona (Italy)
2. Advanced Materials and Liquid Crystal Institute and Department of Physics, Kent State University, Kent, Ohio 44242, USA

**Abstract**

Heliconical cholesteric liquid crystals are expected to be more sensitive to torque induced by light field since their structure allows both bend and twist in molecular orientations, differently from the conventional cholesterics in which only twist deformation is involved requiring much higher fields. We report here a demonstration of tuning the helical pitch in heliconical cholesterics induced by an optical torque. Experimental observations are in agreement with expectations of the classical theory extended to include the effect of the optical field. A dual control of the helical pitch is achieved including both the low frequency electric field applied along the helix axis and the optical field orthogonal to it.

The effects of applied fields on the helical structure of cholesteric liquid crystals have been a subject of extended investigation since the very beginning of research on liquid crystals. In cholesterics due to the chiral nature of the molecules the director **n** twists in space describing a helix, where **n** is always orthogonal to the helix axis[1]. It is well known that this self-assembled periodic structure gives rise to selective Bragg reflection and imposes a strong rotatory power on the incident light[2-4]. The effect of an applied electric field is strongly dependent on its direction with respect to the helix axis and on the elastic constants of the material. The problem has been theoretically solved for an electrostatic (or low frequency) electric field by Meyer[5] and De Gennes[6] in 1968, for materials with positive dielectric anisotropy ($\Delta\epsilon = \epsilon_\parallel - \epsilon_\perp > 0$). An applied field perpendicular to the helix axis was expected to unwind the helix thus increasing the pitch length and shifting the Bragg reflection to the red side of the spectrum. This effect was demonstrated a few years later by Khan[7]. Note that in this geometry, the single-mode periodic modulation of the structure is destroyed by the field, since the director forms wide regions where it is parallel to the field, separated by sharp domain walls where it rotates by 180°.

When the applied field is parallel to the helix axis the perturbation involves both bend and twist deformations therefore the resulting effect depends on the ratio of the corresponding elastic constants $K_3$ and $K_2$, $k = K_3/K_2$. When $k > 1$ the helical axis rotates by $\pi/2$ above a threshold field, while if $k < 1$ a conical deformation may occur, in which the director twists and bends around the heliconical axis. The principal difference between this state and the conventional cholesteric state is that the director is not perpendicular to the helicoidal axis, but makes with it an angle 0<θ<30°. An important advantage of this structure is that the field, applied parallel to the heliconical axis, causes a continuous change of the pitch (which decreases as the field becomes stronger), but does not modify the single-mode character of the periodic structure. While the $\pi/2$ rotation of the helical structure has been observed since the early stage of these investigations[7], only recently the occurrence of the conical deformation has been demonstrated[8-10] and a wide-range tuning of the helical pitch by the applied electric field and by the temperature has been achieved.

Concerning the effects of the optical field, just after the discovery of the Giant Optical Nonlinearity (GON) due to optical reorientation in nematics, the possibility to get an all optical control of the Bragg reflection of cholesterics was considered by using a light beam to induce an optical torque on the molecular director. However, several calculations have pointed out the need of light intensities 3-4 orders of magnitude higher than in nematics to induce an observable effect[11], therefore no clear evidence of light-induced helix unwinding or helix rotation have been provided so far.

The new heliconical phase gives a new opportunity to observe the effect of a light beam on the helical structure because, as already outlined, the interaction with a field involves both twist and bend deformations. While light induced twist is hard to observe due to the high intensity required[12], light induced bend corresponds to the GON effect in nematics, thus it is expected to produce an optical torque comparable to the one originated by the low frequency field and competing with it to determine both the conical angle of the structure and its pitch.

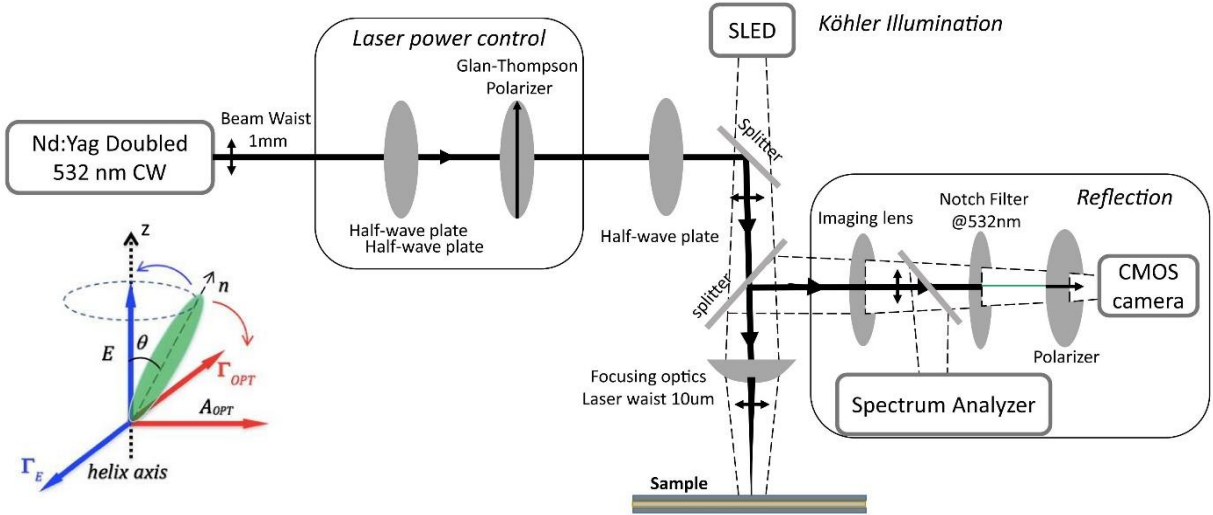

Fig.1 Experimental apparatus. Inset: sketch of the static (E) and optical (A$_{opt}$) fields and of the corresponding torques.

Here we analyze the effect of light on the oblique heliconical cholesterics (Ch-OH) and demonstrate that the helicoidal structure can be affected by an optical field orthogonal to the helix axis. We show that the wavelength of the reflected light can be tuned from green to infrared by changing the power of the incident light. By varying the applied low frequency electric field we demonstrate that tuning of the helical pitch is due to the additional optical torque acted on the molecular director by the light beam. By including in the Meyer's theory the effect of the optical field we find a good agreement with the experimental data.

The free energy density including the contribution of the optical field $\mathbf{E}_{OPT}$ becomes:

$$f = \frac{1}{2}K_1(\nabla \cdot \mathbf{n})^2 + \frac{1}{2}K_2(\mathbf{n} \cdot \nabla \times \mathbf{n} - q_0)^2 + \frac{1}{2}K_3(\mathbf{n} \times \nabla \times \mathbf{n})^2 - \frac{1}{2}\epsilon_0 \Delta\epsilon (\mathbf{n} \cdot \mathbf{E})^2$$
$$- \frac{1}{4}\epsilon_0 \Delta\epsilon_{OPT}(\mathbf{n} \cdot \mathbf{E}_{OPT})^2 \qquad (1)$$

where $K_1$ is the splay elastic constant, $q_0 = 2\pi/P_0$, $P_0$ the helix pitch of the unperturbed cholesteric, $\Delta\epsilon = \epsilon_\parallel - \epsilon_\perp$ at low frequency and $\Delta\epsilon_{OPT} = \epsilon_\parallel^{OPT} - \epsilon_\perp^{OPT}$ at optical frequency.

Equation (1) is calculated by considering the molecular director oriented along the oblique helicoid, $\mathbf{n} = (sin\theta cos\phi, sin\theta sin\phi, cos\theta)$, with the cone angle $\theta > 0$ and the azimuthal rotation angle $\phi = (2\pi/P)z$. The low frequency field is applied along the helix axis direction: $\mathbf{E} = E\hat{\mathbf{z}}$. The optical field is polarized in a plane orthogonal to the helix axis, $\mathbf{E}_{OPT} = E_x\hat{\mathbf{x}} + E_y\hat{\mathbf{y}}$.

If we neglect any light induced twist effect, the effective optical torque acts only on bend deformation. Under this approximation the actual light polarization in the medium is not relevant as far as any longitudinal component of the optical field is negligible. The equilibrium conical angle $\theta$ satisfies the following equation:

$$\frac{\epsilon_0\Delta\epsilon E^2 - \frac{1}{2}\epsilon_0\Delta\epsilon_{OPT}A_{OPT}^2}{K_2^2 q_0^2}[(K_2-K_3)^2\sin^4\theta + 2K_3(K_2-K_3)\sin^2\theta + K_3^2] - K_3 = 0 \quad (2)$$

where $A_{OPT}^2 = (E_x\cos\phi + E_y\sin\phi)^2$ is the square modulus of the effective optical field producing the torque. The competing action of the two applied fields is sketched in the inset of Fig.1, where they are represented with the corresponding torques $\Gamma_E$ and $\Gamma_{OPT}$.

The condition $\sin\theta = 0$ leads to the critical field for complete unwinding:

$$E'_{NC} = \sqrt{E_{NC}^2 + \frac{1}{2}\frac{\Delta\epsilon_{OPT}}{\Delta\epsilon}A_{OPT}^2}, \quad (3)$$

while the usual definition of the critical field in absence of optical pumping is given by $E_{NC} = \frac{2\pi}{P_0}\frac{K_2}{\sqrt{\epsilon_0\Delta\epsilon K_3}}$. The meaning of Eq.(3) is clear: the effect of light is to increase the static field needed to unwind the structure to the nematic state, because the torques of the static and of the optical fields oppose to each other.

At the same time, Eq.(2) suggests that the effective static field along the helix axis is reduced due to the competing action of the optical torque on the molecular director:

$$E_{eff}^2 = E^2 - \frac{1}{2}\frac{\Delta\epsilon_{OPT}}{\Delta\epsilon}A_{OPT}^2. \quad (4)$$

As a consequence, the dependence of the helical pitch on the applied fields, which obeys the equation $P = \frac{2\pi}{E}\sqrt{\frac{K_3}{\epsilon_0\Delta\epsilon}}$ when only the electric field is present[8], is now given by:

$$P = 2\pi\sqrt{\frac{K_3}{\epsilon_0\Delta\epsilon(E^2 - \frac{1}{2}\frac{\Delta\epsilon_{OPT}}{\Delta\epsilon}A_{OPT}^2)}} \quad (5)$$

The pitch increases with the optical field due to the less effective action of the low frequency field. Eq.(5) results in the condition: $E^2 \geq \frac{1}{2}\frac{\Delta\epsilon_{OPT}}{\Delta\epsilon}A_{OPT}^2$. However since the minimum necessary for the onset of the heliconical structure when only the electric field is present is[8] $E_{N^*C} = E_{NC}\frac{k[2+\sqrt{2(1-k)}]}{1+k}$, it means that in presence of the optical field this becomes:

$$E'_{N^*C} = \sqrt{E_{N^*C}^2 + \frac{1}{2}\frac{\Delta\epsilon_{OPT}}{\Delta\epsilon}A_{OPT}^2} \quad (6)$$

We obtained the Ch-OH by mixing a dimeric LC 1'',7'' - bis(4-cyanobiphenyl-4'-yl) heptane (CB7CB), rod-like mesogen pentylcyanobiphenyl (5CB) (Merck) and left-handed chiral dopant S811 (Merck), in weight proportion 5CB:CB7CB:S811 = 44.7:51.1:4.2. At $61.5°C$, the cholesteric melts into an isotropic fluid, and at $26.7°C$ it transforms into the chiral analog of the twist-bend nematic phase[13-15]. The mixture was sandwiched between two conductive glasses treated to get planar alignment of the director. Mylar spacers 20 µm thick were used to control cell thickness. The cell was filled keeping the material in the isotropic phase and then slowly cooled down to room temperature.

The used mixture exhibits the behavior typical of Ch-OH[8-10]: above a sufficiently strong electric field, the material is switched into a uniform nematic with the director parallel to the field, giving rise to a state which appears dark when viewed between crossed linear polarizers. By decreasing the external field, the LC shows a sequence of changing wavelength of reflection, from UV to visible blue, then green, orange, red, and, finally, near IR. By further decreasing the field, one obtains a light scattering texture. All data reported here were obtained at *27.5° C*.

The effect of light on the cell was investigated using the inverted microscope apparatus shown in Fig. 1. The pump beam is originated by a CW frequency doubled Nd:YAG laser (λ = 532 nm), focused by a 5 cm lens in the sample volume. The probe beam is the white light of a SLED source, focused by the same optics used for the pump. The used set up allows us to monitor both the reflection spectra and the sample appearance in reflection mode. A half wave plate combined with a Glan-Thomson polarizer were used to vary the power of the pump beam. Cell temperature was controlled by means of a Peltier temperature controller with an accuracy of 0.1°C.

Irradiation with light polarized in the plane of the cell with no electric field applied does not produce any selective reflection of light, indicating that, in the used experimental conditions, the Ch-OH structure cannot be induced by the optical field alone. This behavior is in agreement with previous observations stating that the Ch-OH phase requires a non-vanishing component of the field along the helix axis[8,9]. Tilting the cell does not change the situation, which can be an indication of a too small axial component of the field.

The behavior under the combined action of the electric and optical fields is different. Application of a low frequency field (in the following "static") along the helix with the amplitude above a threshold value $E_{NC} = 4.5 \frac{V}{\mu m}$, unwinds the helix. Lower fields allow the existence of the heliconical structure as already described, that is evidenced by the appearance of different colors in the reflection pattern. The helical pitch decreases as the field increases, as shown in the first line of Fig.2, where the black pattern corresponds to Bragg reflection in the UV region. The color-tuning effect of a pump beam illuminating the same cell under the same static field is shown in the second line; the beam is linearly polarized in the plane of the cell.

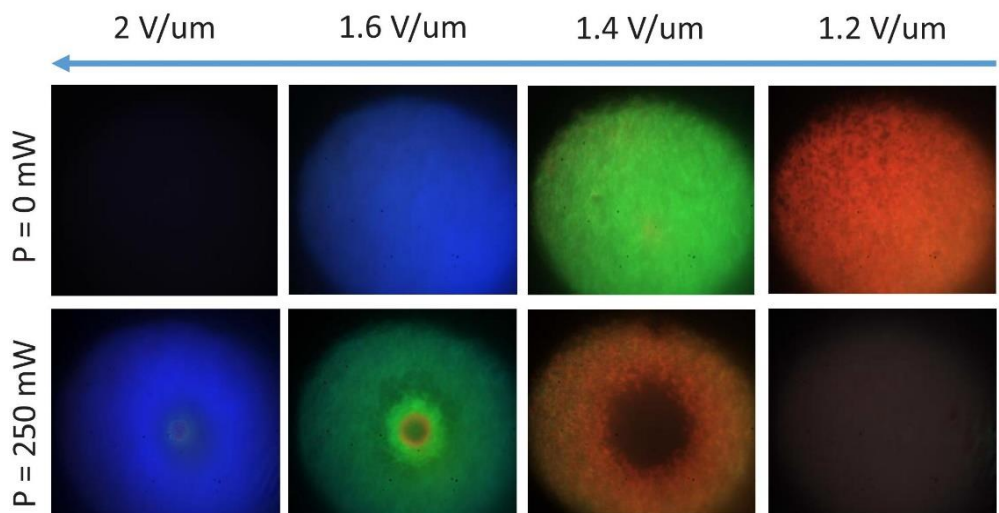

Fig.2 Sample appearance in reflection mode under the action of an increasing static electric field in the absence (left column) and in the presence (right column) of an optical field. Light power: 250 mW, beam waist: 10 µm.

The effect of pitch elongation induced by the light field is made evident by the observed colors: the impinging light shifts the sample reflection to longer wavelengths. In these pictures it is also evident a strong effect due to nonlocality since the image corresponds to an area larger than the laser waist on the sample (about 10 µm). The dark spots in the center of the images (wider as the static field decreases) correspond to reflection in the near IR region. This effect shows a very weak dependence on pump polarization.

As outlined, laser irradiation induces an increase of the pitch for each applied voltage. The same effect is observed by increasing temperature[10] and in our sample we measured a shift of the reflection peak from 430 to 610 nm with a change of only 3°C (from 27 to 30 °C) for a fixed static field of 2.2 V/µm. Such a high sensitivity to temperature variations can involve light-induced heating due to absorption that contributes to the observed color changes. In order to understand the origin of the light-induced effect a different experiment was performed by keeping the static field at a value over 4.5 V/µm, where the oblique helical structure is unwound and the director is aligned in a uniaxial fashion parallel to the static field. Under this condition the sample appears dark when viewed between crossed polarizers; this homeotropic state is preserved as the temperature rises. Since the oblique helicoid is absent, the heating cannot modify the Bragg reflection.

By choosing $E > E_{NC}$ at any fixed temperature, irradiation with light causes the colored pattern to reappear, that is light restores the colored appearance of the sample originating in Bragg reflection, as shown in Fig. 3 for a pump power P = 300 mW. A few seconds of light irradiation produce selective reflection of light. Similarly to Fig.2, the irradiated area is much smaller than the colored one in the image and the wavelength of the selective reflection decreases as one moves from the center of the beam to the periphery and to the not irradiated region where nonlocal reorientation can take place. This observation eliminates the possibility of the light-induced heating as the cause of the spectral changes and is consistent with optical reorientation triggered by the optical torque that competes with the static field.

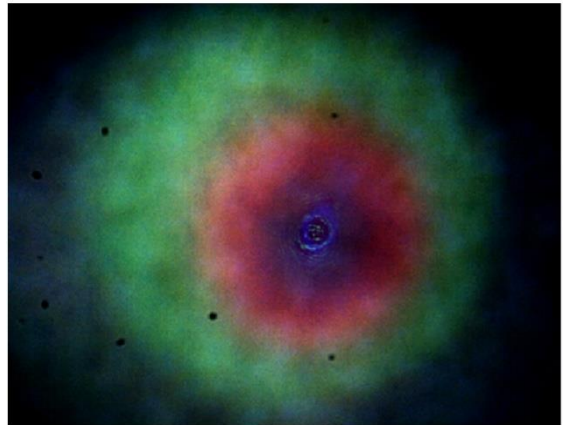

Fig.3 Sample appearance in reflection mode under the combined action of light and of a static electric field higher than the critical one required to unwound the helix.

In order to demonstrate the occurrence of the optical reorientation effect, the following simple measurement can be carried out. In the presence of an optical field, according to Eq.(3), the critical voltage necessary to produce the unwound nematic structure increases:

$$V'_{NC} = \sqrt{V_{NC}^2 + \frac{\Delta\epsilon_{OPT}}{\Delta\epsilon} d^2 \frac{\epsilon_r^2 I}{\epsilon_0 c n_{av}}}, \qquad (7)$$

where we use the following relationship between the effective optical field and the intensity $A^2 = \frac{2I}{\epsilon_0 c n_{av}}$ ($n_{av}$ is an average refractive index).

Measurements have been performed at different pump powers, monitoring how the critical voltage depends on light intensity. Critical voltage was determined at the occurrence of a dark texture between crossed polarizers.

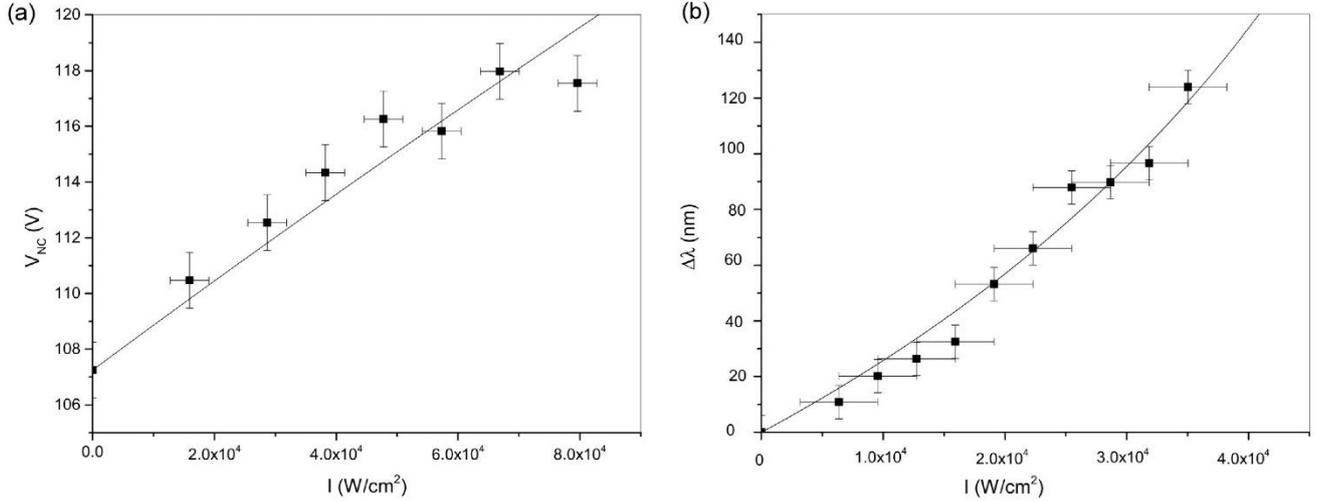

Fig.4 a) Critical voltage necessary to unwound the structure vs light intensity. Continuous line is a best fit with eq. (7); b) Reflected wavelength variation $\Delta\lambda$ vs light intensity and best fit with eq. (10). External electric field: 2.28 V/μm.

Data are reported in Fig. 4a with a best fit using eq.(7). The experimental data are in a good agreement with the theory. The fitting parameter $B$ defined by the relation $V'_{NC} = \sqrt{V_{NC}^2 + B\,I}$ has the value $B = 3.6 \cdot 10^{-6}\ V^2\ m^2/W$ which is very close to the theoretical one $\frac{\Delta\epsilon_{OPT}}{\Delta\epsilon}\frac{\epsilon_r^2}{\epsilon_0 c n_{av}} d^2 = 2 \cdot 10^{-6}\ V^2\ m^2/W$ obtained using the values of the material parameters of a very similar mixture where they have been measured ($\Delta\varepsilon_{opt} \cong 0.64$, $\Delta\varepsilon \cong 8$, $\varepsilon_r = \varepsilon_{//} \cong 14$, $n_{av} \cong 1.6$). According to this result, we expect that at any value of static field that stabilizes a heliconical structure, light illumination increases the pitch, which is evidenced by the red-shift of the Bragg peak. To verify this hypothesis we measured the reflected wavelength as a function of the light intensity, for a fixed value of the static field. The latter was kept at 2.28 V/μm which corresponds to an "unperturbed" reflected wavelength $\lambda_0$ = 427 nm. Results are shown in Fig. 4b, where the solid line is a fit with eq. (9).

By introducing the light intensity eq.(5) becomes:

$$P = 2\pi \sqrt{\frac{K_3}{\epsilon_0 \Delta\epsilon (E^2 - \frac{\Delta\epsilon_{OPT}}{\Delta\epsilon}\frac{I}{\epsilon_0 c n_{av}})}} \qquad (8)$$

The shift of the reflected wavelength observed in Fig. 4b is determined by the pitch variation: $\Delta\lambda = n_{av}[P(I) - P(0)]$. By substituting eq. (8) in this expression one gets:

$$\Delta\lambda = n_{av}\left[2\pi\sqrt{\frac{K_3}{\epsilon_0\Delta\epsilon(E^2 - \frac{\Delta\epsilon_{OPT}}{\Delta\epsilon}\frac{I}{\epsilon_0 c n_{av}})}} - P(0)\right] \quad (9)$$

which has been used to fit the curve in Fig.4b.

The agreement is again quite good. From the parameter $A = 2\pi\sqrt{\frac{K_3}{\epsilon_0\Delta\epsilon}}$ we can estimate the bend elastic constant of the used mixture: $K_3 \cong 0.6\, pN$, which is in good agreement with the values found in similar mixtures exhibiting the oblique heliconical phase at temperature close to the transition to the twist-bend nematic phase[8,16], thus strengthening our confidence in the good quality of the fit. The parameter $\frac{\Delta\epsilon_{OPT}}{\Delta\epsilon}\frac{1}{\epsilon_0 c n_{av}}$ turns out to be higher than the one calculated but still of the right order of magnitude ($0.29 \cdot 10^2$ vs the calculated $0.19 \cdot 10^2$ $V^2/W$).

Figure 5 shows how the minimum and the maximum static critical fields required to observe the heliconical cholesteric phase, namely $E'_{N*C}$ and $E'_{NC}$, depend on the light intensity. They both increase by the same amount, in agreement with equations (3) and (6), showing that light shifts upward the static field range able to stabilize the heliconical structure, in a way very different from temperature rise that has been reported to significantly broaden this range. Therefore these data are an additional proof of the light-controlled spectral response.

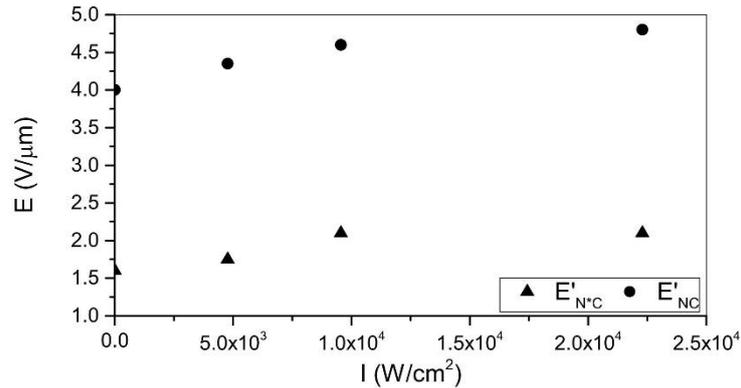

Figure 6. Minimum (full triangles) and maximum (full circles) critical static fields necessary to observe the heliconical phase, vs light intensity.

We reported the first observation of optical control of the helical pitch in Ch-OH and the optical tuning of their structural colors caused by selective Bragg reflection. We demonstrated that the Ch-OH structure can be affected by an optical field orthogonal to the helix axis. We showed that the wavelength of the reflected light can be tuned from green to IR by changing the power of the incident light.

The effect is due to the optical torque acted on the molecular director by the light beam, that is opposite to the one caused by the static field, as demonstrated by measuring how the intensity of impinging light affects the effective threshold field required to totally unwind the helix and the wavelength of the Bragg reflection. By supplementing the Meyer-de Gennes theory with the effect of the optical field, we find a good agreement with the experimental data.

**Acknowledgements**
L.L, F.S. and G.N. would like to thank Victor Reshetnyak and Tommaso Bellini for useful discussions. The work of O.S.I. and O.D.L has been supported by NSF grant ECCS-1906104.